\input harvmac
%\draft
%-------------------------
% This paper uses harvmac
%-------------------------
\overfullrule=0pt
\def\Title#1#2{\rightline{#1}\ifx\answ\bigans\nopagenumbers\pageno0\vskip1in
\else\pageno1\vskip.8in\fi \centerline{\titlefont #2}\vskip .5in}

\font\ticp=cmcsc10
%
%-------------------
%  definitions
%-------------------

%
\def\half{{1\over2}}

\def\e{\epsilon}

%
%-------------------
% references
%
\lref\cvetd{M. Cvetic and D. Youm, hep-th/9507090.}
\lref\chrs{D. Christodolou, Phys. Rev. Lett. {\bf 25}, (1970) 1596;
D. Christodolou and R. Ruffini, Phys. Rev. {\bf D4}, (1971) 3552.}
\lref\cart{B. Carter, Nature {\bf 238} (1972) 71.}
\lref\penr{R. Penrose and R. Floyd, Nature {\bf 229} (1971) 77.}
\lref\hawka{S. Hawking, Phys. Rev. Lett. {\bf 26}, (1971) 1344.}
\lref\sussc{L.~Susskind,  Phys. Rev. Lett. {\bf 71}, (1993) 2367;
L.~Susskind and L.~Thorlacius, Phys. Rev. {\bf D49} (1994) 966;
L.~Susskind, ibid.  6606.}
\lref\polc{J. Dai, R. Leigh and J. Polchinski, Mod. Phys.
Lett. {\bf A4} (1989) 2073.}
\lref\hrva{P. Horava, Phys. Lett. {\bf B231} (1989) 251.}
\lref\cakl{C. Callan and I. Klebanov, hep-th/9511173.}
\lref\prskll{J. Preskill, P. Schwarz, A. Shapere, S. Trivedi and
F. Wilczek, Mod. Phys. Lett. {\bf A6} (1991) 2353. }
\lref\sbg{S. Giddings, Phys. Rev {\bf D49} (1994) 4078.}
\lref\cghs{C. Callan, S. Giddings, J. Harvey, and A. Strominger,
Phys. Rev. {\bf D45} (1992) R1005.}
\lref\bhole{G. Horowitz and A. Strominger,
Nucl. Phys. {\bf B360} (1991) 197.}
\lref\bekb{J. Bekenstein, Phys. Rev {\bf D12} (1975) 3077.}
\lref\hawkb{S. Hawking, Phys. Rev {\bf D13} (1976) 191.}
\lref\wilc{P. Kraus and F. Wilczek, hep-th/9411219, Nucl. Phys.
{\bf B433} (1995) 403. }
\lref\intrp{G. Gibbons and P. Townsend, Phys. Rev. Lett.
{\bf 71} (1993) 3754.}
\lref\gmrn{G. Gibbons, Nucl. Phys. {\bf B207} (1982) 337;
G. Gibbons and K. Maeda Nucl. Phys. {\bf B298} (1988) 741.}
\lref\bch{J. Bardeen, B. Carter and S. Hawking,
Comm. Math. Phys. {\bf 31} (1973) 161.}
\lref\mdr{W. Zurek and K. Thorne, Phys. Rev. Lett. {\bf 54}, (1985) 2171.}
\lref\stas{A.~Strominger and S.~Trivedi,  Phys.~Rev. {\bf D48}
 (1993) 5778.}
\lref\jpas{J.~Polchinski and A.~Strominger,
hep-th/9407008, Phys. Rev. {\bf D50} (1994) 7403.}
\lref\send{A. Sen, hep-th/9510229, hep-th/9511026}
\lref\cvet{M. Cvetic and A. Tseytlin, hep-th/9512031.}
\lref\kall{R. Kallosh, A. Linde, T. Ortin, A. Peet and
A. van Proeyen, Phys. Rev. {\bf D46} (1992) 5278.}
\lref\lawi{F. Larsen and F. Wilczek, hep-th/9511064.}
\lref\bek{J. Bekenstein, Lett. Nuov. Cimento {\bf 4} (1972) 737,
Phys. Rev. {\bf D7} (1973) 2333, Phys. Rev. {\bf D9} (1974) 3292.}
\lref\hawk{S. Hawking, Nature {\bf 248} (1974) 30, Comm. Math. Phys.
{\bf 43} 1975.}
\lref\sen{A. Sen, hep-th/9504147.}
\lref\suss{L. Susskind, hep-th/9309145.}
\lref\sug{L. Susskind and J. Uglum, hep-th/9401070,Phys. Rev. {\bf D50}
 (1994) 2700.}
\lref\peet{A. Peet, hep-th/9506200.}
\lref\tei{C. Teitelboim, hep-th/9510180.}
\lref\carl{S. Carlip, gr-qc/9509024. }
\lref\thoo{G. 'tHooft, Nucl. Phys. {\bf B335} (1990) 138;
Phys. Scr. {\bf T36} (1991) 247.}
\lref\fks{S. Ferrara, R. Kallosh and A. Strominger, hep-th/9508072,
Phys. Rev. {\bf D 52}, (1995) 5412 .}
%-------------------
% title page
%-------------------
%
\Title{\vbox{\baselineskip12pt
\hbox{HUTP-96/A002}\hbox{RU-96-01}
\hbox{hep-th/9601029}}}
{\vbox{
\centerline{Microscopic Origin}
\centerline{of the}\centerline {Bekenstein-Hawking Entropy}}}

\centerline{{\ticp Andrew Strominger}}

\vskip.1in
\centerline{\sl Department of Physics}
\centerline{\sl University of California}
\centerline{\sl Santa Barbara, CA 93106-9530}

\smallskip
\centerline{and}
\smallskip
\centerline{{\ticp Cumrun Vafa}}

\vskip.1in
\centerline{\sl Lyman Laboratory of Physics}
\centerline{\sl Harvard University}
\centerline{\sl Cambridge, MA 02138}

\bigskip
\centerline{\bf Abstract}
The Bekenstein-Hawking area-entropy relation $S_{BH}=A/4$ is derived for
a class of five-dimensional extremal black holes in string theory
by counting the degeneracy of BPS soliton bound states.
\Date{}
%
%----------------------
% Body of Paper
%----------------------
\newsec{Introduction}
In the early seventies a sharp and beautiful analogy was discovered
between the laws of black hole dynamics
and the laws of thermodynamics \refs{\chrs\penr\hawka\bek\cart
\bch - \hawk}.
In particular the Bekenstein-Hawking entropy - one quarter the area
of the event horizon - behaves in every way like a thermodynamic
entropy. A missing link in this circle of ideas is a precise
statistical mechanical interpretation of black hole entropy.
One would like to
derive the Bekenstein-Hawking entropy - including the numerical factor -
by counting black hole microstates. The laws of black hole
dynamics could then be identified with - and not just be
analogous to - the laws of thermodynamics.

In this paper progress in this direction is reported.  We consider phases
of string theory with five noncompact dimensions and $N=4$ 
supersymmetry\foot{Analogous results follow for $N=8$ as indicated below.},
({\it e.g.} type II string theory on $K3\times S^1$ or heterotic string
theory on $T^5$). Black holes in these theories can carry
both an axion charge $Q_H$ and an electric charge $Q_F$\foot{$Q_F\in
\Gamma^{5,21}$
where $\Gamma^{5,21}$ is the Narain lattice of heterotic strings
compactified down to 5 dimensions and $Q_F^2=Q_R^2-
Q_L^2$.}. Extremal black holes with either $Q_H=0$ (fundamental heterotic
string states)
or $Q_F=0$
(but not both) have degenerate horizons with zero area.  We
accordingly look for BPS saturated states
- {\it i.e.} extremal black holes - for which both
$Q_F$ and $Q_H$ are non-vanishing.
Such BPS states preserve
only $1/4$ of the $N=4$ supersymmetry.
They may be viewed as bound states of
minimally-charged BPS solitons, and their
exact degeneracy as a function of
$Q_F$ and $Q_H$ can be topologically computed by
counting soliton bound states.
In particular we
show that the leading degeneracy for
the logarithm of the bound-state degeneracy
for large $Q_H$ and fixed $Q_F$ is
given by\foot{Given the $O(21,5)$ invariance
of the theory one expects that
the bound-state degeneracy of these BPS
solitons be a functions of $Q_F^2$ and $Q_H$.}
\eqn\bsd{S_{stat}=2\pi \sqrt{Q_H({1\over 2}Q_F^2+1) }.}
On the other hand we will find that the Bekenstein-Hawking entropy as
determined from the low-energy effective action is
\eqn\bhsd{S_{BH}=2\pi \sqrt{Q_HQ_F^2 \over 2}.}
in agreement with \bsd\ for large charges.

The five-dimensional problem is considered here because it
seems to be the simplest non-trivial case.
We expect that similar calculations will reproduce $S_{BH}$ for other types
of black holes in string theory. Previous attempts at a microscopic
derivation of $S_{BH}$ include
\refs{\bekb\hawkb \mdr\thoo\suss\sug\tei\sen\carl\lawi -\cvet}.

In section 2 we present the black hole solutions and compute the area
of their event horizons. In section 3 the bound state degeneracy is
asymptotically computed by
relating it to the elliptic genus of a certain two-dimensional
sigma model. We conclude with discussion in section 4.

\newsec{A Class of Five Dimensional Extremal Black Holes}

The low energy action for type II string theory compactified
on $K3\times S^1$
contains the terms
\eqn\fds{{1\over 16 \pi}
\int d^5x \sqrt{-\tilde g}\bigl(e^{-2\phi}\bigl( R+4(\nabla \phi )^2
-{1 \over 4}\tilde H^2\bigr)-{1 \over 4} F^2\bigr)}
in the string frame. We adopt conventions in
which $\alpha^\prime =G_N=1$. $F$ is a RR 2-form field strength
(associated with the
right-moving current algebra in the dual heterotic picture) and $\tilde H$
is a 2-form
axion field strength arising from the NS-NS 3-form with
one compnent tangent to the $S^1$. We work on a submanifold of the Narain
moduli space for $K3\times S^1$ on 
which nonzero $F$ does not require nonconstant moduli.
In the Einstein frame ($g=e^{-4\phi/3} \tilde g$) \fds\ becomes
\eqn\fdde{{1\over 16 \pi}\int d^5x \sqrt{- g}\bigl(
R-{4 \over 3}(\nabla \phi )^2-{e^{-4\phi/3} \over 4}\tilde H^2-{e^{2\phi/3}
 \over 4}F^2\bigr).}
A black hole can carry electric charge with respect to both
$F$ and $\tilde H$,
\eqn\qhd{\eqalign{Q_H&\equiv {1/4\pi^2}\int_{S^3}* e^{-4\phi/3}\tilde H ,\cr
 Q_F&\equiv {1/16\pi}\int_{S^3} *e^{2\phi/3}F .\cr}}
For the spherically symmetric configurations that we consider
this implies
\eqn\qqd{\eqalign{* e^{-4\phi/3}\tilde H&= 2Q_H\e_3  ,\cr
*e^{2\phi/3}F & = {8 Q_F \over \pi}\e_3 ,\cr}}
where $\e_3$ is the volume element on the unit $S^3$.
We have chosen our conventions so that $Q_H$ and ${1\over 2}Q_F^2$
are integers\foot{
$Q_F^2=2$ for a minimally-charged perturbative
heterotic string state, while $Q_H=1$ for a minimally-charged
heterotic fivebrane which wraps $T^5$ once.}. $O(21,5)$ invariance of the
full lagrangian (which includes 26 gauge fields) implies that
all of the following
formulae remain valid with the replacement $Q^2_F=Q^2_R-Q^2_L$.

An extremal black hole carrying both types of charges
can have an event horizon with nonzero area.
The near-horizon geometry will be the five-dimensional
$AdS_2\times S^3$ charged Robinson-Berttoti universe with
constant dilaton $\phi=\phi_h$.
The constant value $\phi_h$ is determined in terms of the
charges by the dilaton  equation of motion
\eqn\deom{16\nabla^2 \phi +2 e^{-4\phi/3}\tilde H^2-
e^{2\phi/3}F^2=0.}
Substituting $\phi=\phi_h$ and \qqd\ this implies
\eqn\std{e^{2\phi_h}=
\half ({4Q_F \over\pi Q_H })^2.}
Note that the type II closed string coupling at the horizon
is weak when the ratio $Q_F/Q_H$ is small.
If the asymptotic value of $\phi_\infty$ of the dilaton is tuned
to coincide with the special value determined by
\std\ then the dilaton is everywhere
constant and
Einstein's equation becomes
\eqn\enst{\eqalign{R_{ab}&=
3({8Q_HQ_F^2 \over \pi^2})^{2/3}\bigl(\e_{3acd}{\e_{3b}}^{cd}-g_{ab}\bigr) ,\cr
\phi&=\phi_h.}}
This is just the equation for  $d=5$ Reissner-Nordstrom
with charge $\sqrt{3}({8Q_HQ_F^2 \over \pi^2})^{1/3}$. The
extremal solution
can be found for example in \gmrn :
\eqn\exmet{
ds^2=-\bigl(1-({r_0 \over r})^2\bigr)^2dt^2+\bigl(1-({r_0 \over r})^2
\bigr)^{-2}dr^2
+r^2d\Omega^2_3,}
where
\eqn\erk{r_0=({8Q_HQ_F^2\over \pi^2})^{1/6}.}
$S^1$ reduction of this solution to $d=4$ gives the
dyonic solution discussed in reference
\refs{\kall}, where it is further shown that
the resulting configurations are annihilated by
one quarter of the supersymmetries (This is also
evident from the stringy description given below).

The Einstein-frame area of the extremal black hole horizon is
given by the volume of the $S^3$:
\eqn\aria{Area=8 \pi \sqrt{{Q_HQ_F^2 \over 2}}.}
The Bekenstein-Hawking entropy is
\eqn\ent{S_{BH}=2\pi \sqrt{{Q_HQ_F^2 \over 2}}.}

Even if the asymptotic value of the dilaton $\phi_\infty \neq \phi_h$,
the near-horizon Robinson-Berttoti geometry is still
constrained to obey \std\ and \enst. Hence as the asymptotic
value of the fields are adiabatically changed, the near horizon geometry is
unaltered. This type of behavior has  been noticed previously in families of
exact solutions with generic asymptotic moduli (see for example
\refs{\fks,\lawi,\cvetd,\cvet}) and can be intuitively understood \intrp\ by
viewing
black holes as solitons which interpolate between maximally symmetric
vacua at infinity and the horizon
(in our case the $d=5$ Robinson-Berttoti vacuum). In conclusion the
dilaton-independent
relation \ent\ is valid even when $\phi_\infty \neq \phi_h$.

The action \fds\ as well as the entropy \ent\ receives
corrections from both string
loop and sigma model perturbation theory.
$N=4$ nonrenormalization theorems ensure that
there are no corrections to the lowest dimension terms
exhibited in \fds, but higher dimension terms will be corrected in general.
Type II string loop corrections are suppressed by powers of
$g_{II} \sim {Q_F/Q_H}$. Sigma model corrections are
suppressed by inverse powers of the string-frame Schwarzchild
radius, which is $\tilde r_{0II} \sim \sqrt {Q_F^2/Q_H}$.\foot{For the dual
heterotic theory, $g_h^2 \sim (Q_H/Q_F)$ and $\tilde r_{0h} \sim \sqrt Q_H$.}
Hence validity of \ent\ in the type II  theory
requires that both $Q_H$ and $Q_F$ are large. 
String dualities of various kinds 
might be used to extend the range of validity 
of \ent. 

\newsec{Counting of Microscopic BPS States}

\lref\witb{E. Witten, hep-th/9510135.}\lref\vgas{C. Vafa,
hep-th/9511088.}\lref\bsv{M. Bershadsky, V. Sadov and C. Vafa,
hep-th/9511222.}\lref\vins{C. Vafa, hep-th/9512078.}

The counting of microscopic BPS states has become
possible for type II string compactifications
thanks to recent progress in understanding non-perturbative string theory.
Of particular importance is the beautiful
identification of D-branes \refs{\polc,\hrva} as the source of BPS
states carrying $Q_F$ Ramond-Ramond charge
\ref\pol{J. Polchinski, hep-th/9510017.}\ and the relation between counting
bound states of D-branes and specific questions
in certain quantum field theories on the D-brane worldvolume \refs{
\witb ,\send, \vgas ,\bsv ,\vins }.

Consider type IIB string theory compactified on $K3\times S^1$.  Type
IIB string theory has $p$-brane solitonic
states for odd values of $p$ \bhole.
We consider D-branes with $p=1,3,5$ wrapped around $S^1\times C$
where $C$ is a supersymmetric ({\it i.e.} holomorphic)
$0$-, $2$- or $4$-cycle of $K3$.   These states
carry the Ramond-Ramond charge $Q_F$, and $ Q_F^2$ is the
self-intersection number of the collection of cycles \refs{\send,\bsv,
\vgas,\vins}.
It was argued in
\witb\ that BPS states in spacetime which preserve half of the
spacetime supersymmetries correspond to supersymmetric
ground states of the D-brane worldvolume
theory.  This follows directly from the fact that worldvolume supersymmetries
arise as the projection of unbroken spacetime supersymmetries.
This observation was generalized in \vins\ to spacetime BPS states,
which preserve fewer spacetime supersymmetries.  The
corresponding states of the D-brane worldvolume theory correspond to
worlvolume BPS states, which preserve fewer worldvolume supersymmetries
than the worldvolume ground states.
Since we are interested in states which preserve $1/4$ of the
spacetime supersymmetries
we should count BPS states which preserve $1/2$ of the
supersymmetries of the relevant D-brane worldvolume theory.
Let us first consider
a limit in which the worldvolume theory of the D-brane
simplifies.  Consider the limit in which the $K3$ is small
compared to the size of the circle $S^1$.  In this limit
we get an effectively two dimensional worldvolume theory
on $S^1\times R$ (where $R$ corresponds to time).  Based on
the expected ground state degeneracies of the corresponding
effective theory\foot{According to string duality this is expected to be
the same as the oscillator
degeneracy of bosonic strings.}
it was conjectured
in \vgas\ that this theory is a supersymmetric
sigma model whose target space is the symmetric product of
$\half Q_F^2 +1$ copies
of $K3$
\eqn\simo{M={(K3)^{\otimes \big[{1\over 2}Q_F^2+1\big]}\over S_{\big[{1\over
2}Q_F^2+1
\big]}}}
where $S_n$ is the permutation group on $n$ objects.
Subsequently this
was verified \bsv\ at least for cases where $Q_F$ comes from
primitive 2-cycles in $K3$
or from $[\half Q_F^2+1]$ $0$-cycles together with one 4-cycle in $K3$ (for
2-cycles this was rephrased and further checked as
a counting problem for rational curves on $K3$ with double
points \ref\zyau{
S.-T. Yau and E. Zaslow, hep-th/9512121.}).
In the latter case the origin of \simo\ can be simply understood
as the moduli space of $\half Q_F^2+1$ unordered points on $K3$.
The conjecture \simo\ was further verified in \vins\
in cases where there are more than one 4-cycle
and some 0-cycles
(and was connected to the strong coupling test of Olive-Montonen
duality on $K3$ \ref\vwit{C. Vafa and E. Witten, Nucl. Phys. {\bf B431}
(1994) 3.}).  For the purposes of this paper consideration
of any of these configurations of D-branes suffices\foot{In
fact all we will need is that the dimensions of $M$ grows
as $4(\half Q_F^2+1)$, which is much easier to argue
\refs{\bsv,\vins}.}.

\lref\witel{E. Witten,
Comm. Math. Phys. {\bf 109} (1987) 525.}\lref\sw{A.N.
Schellekens and N.P. Warner, Phys. Lett. {\bf B177} (1986) 317.}

Thus we have to count the BPS states of a supersymmetric sigma model on $M$
if we wish to count states which preserve only $1/4$ of the spacetime
supersymmetries.
But these are precisely those states which are killed by (say) the right-moving
supercharge, with no restrictions on the left-movers.  In other words
we consider RR sector states of this sigma model in their
right-moving vacuum, i.e.,
${\overline L_0}={1\over 2}(H-P)=0$ and arbitrary $L_0={1\over 2}(H+P)$.  The
generating function for the degeneracies of such states is
 bounded by the elliptic genus of the sigma model on $M$ \refs{\witel ,
\sw }\ which is computable for all manifolds $M$.  The actual
number of BPS states may depend on the moduli of $K3$, but
the elliptic genus, which is the appropriate weighted sum (with
$\pm 1$), is moduli independent.  It is tempting
to speculate that the elliptic genus is the more
relevant quantity which appears in physical quantities
(just as was the case considered in \ref\mhar{J. Harvey
and G. Moore, hep-th/9510182.})
but either quantity will give the same leading degeneracy
 as a function of charges \ref\kv{I. Kani and C.Vafa,
 Comm. Math. Phys. {\bf 130} (1990) 529.}, and this distinction is unimportant
for our purposes in this paper.
Note that the eigenvalues
of $L_0$ contributing to the elliptic genus
are restricted to be integers because $L_0=L_0- {\overline L_0}=P$
where $P$ is the momentum operator on the $S^1$.

Before considering the
degeneracy of these states let us see what charges they carry.
In addition to $Q_F$ charge they carry a charge corresponding
to momentum $P$ around $S^1$.  If we go from type IIB to type IIA by
dualizing the $S^1$, these states carry $P$ units of
winding around $S^1$, i.e. they have $P$ units
of electric charge with respect to
$B_{\mu \theta}$ where $\theta$ corresponds to the circle direction and
$\mu$ denotes the five dimensional spacetime indices, i.e. $Q_H=P$.
Thus the BPS states of the D-brane
worldvolume theory we are considering carry precisely the charges
$Q_F$ and $Q_H$ for which the corresponding extremal black hole
solutions were found
in the previous section.

To compare their degeneracy with the Bekenstein-Hawking entropy \ent , which
is expected to be accurate for large $Q_H$ and $Q_F^2$,
all we need to do is to consider the asymptotic degeneracy of the above BPS
states for large $Q_H=P$ and $Q_F^2$ (even though it
is straight forward to compute it for all $Q_H$ and $Q_F^2$ by an orbifold
computation).  For a hyperkahler manifold $M$ of dimension $4k$ we
have a sigma model with central charge $c=6k$.  The left-moving
oscillator degeneracy for unitary conformal theories at level $L_0=n$
goes for $n>>1$ as \ref\car{J.L. Cardy, Nucl. Phys. {\bf B270} 
(1986) 186.}\foot{It would be interesting to understand large $c$ 
corrections to this formula in order to determine the range of validity of 
our estimate.} 
\eqn\degn{d(n,c)\sim {\rm exp}({2\pi \sqrt{nc\over 6}})}
In our case
\eqn\cnr{\eqalign{c&=6({1\over 2}Q_F^2+1), \cr n&=Q_H,}} so we get
for the growth of the elliptic genus, or equivalently the degeneracy of BPS
solitons for $Q_H>>1$
\eqn\ddeg{S_{stat}=\ln d(Q_F,Q_H)\sim  {2\pi \sqrt{Q_H({1\over 2}Q_F^2+1})}.}
This agrees to leading
order with the expected Bekenstein-Hawking entropy \ent\ for
large $Q_F^2$, in which case \ent\ is reliable\foot{This result can also 
be derived for an $N=8$ toroidal type II compactification, in which case 
the same type of D-brane configuration breaks $7/8$ of the supersymmetry. 
The only difference in the derivation is that $K3$ is replaced by $T4$ in 
the symmetric product in \simo. Since the dimension of the resulting 
$M$ is the same this does not affect the growth of the elliptic genus.
We thank A. Sen for discussions on this point. }

\newsec{Discussion}

 In the presence of $N$ D-branes, the open string sector of perturbation
theory involves an expansion in $g_{II}N$
(where $g_{II}$ is the asymptotic value of the type II closed
string coupling)
because holes in the string world
sheet can have $N$ types of
Dirichlet boundary conditions.
The effective value of $N$
for our configurations grows like $\half Q_F^2$.
Hence for small but fixed $g_{II}$ string perturbation theory
will break down for sufficiently large
charge. The correct physical picture of the objects we
discuss really is as a large semiclassical black hole with
an event horizon. The description
as a supersymmetric cycle embedded in $K3$ suffers large
quantum corrections.
It nevertheless can be reliably used to compute the asymptotic
degeneracy of
BPS states because that is a
topological quantity related to the elliptic genus.

The validity of string perturbation theory can be
restored by taking $g_{II}$ to be very
small - smaller than $1/N$. In this case string
perturbation theory is valid, and the
physical picture of the BPS state as a  supersymmetric $K3$
cycle is the correct weakly-coupled description.
For such very small $g_{II}$, the string
length becomes larger than the Schwarzchild radius (equation \erk ).
Hence the
black hole picture will
suffer
large stringy corrections.

So we have a BPS state which at very weak coupling is
described by $p$-branes wrapping supersymmetric $K3$
cycles, but at strong coupling transforms into a hole in
spacetime!
% Lately we have become
%used to the idea that BPS states may have wildly different
%descriptions in different
%parts of moduli space, but this is surely one of the wildest!

We believe that our results will have implications for the
black hole information puzzle. A central theme in studies
of this puzzle over the last several years
has been the problem of low-energy scattering
of ordinary quanta by an extremal black hole \refs{\prskll,\cghs,\stas,\sbg,
\wilc}.
Naively this process proceeds by absorption followed by Hawking
reemission, and so the question of unitarity violation arises.

In principle light might be shed on this puzzle by employing
D-brane technology \refs{\pol,\cakl} to compute the scattering.
However one immediately encounters the above-mentioned
problem that string theory is strongly coupled in the region of interest.
Perhaps a string duality can be used to map it to a weakly-coupled problem.
In any case it is hard to imagine how any calculation based on our D-brane
description of the extremal black hole
could yield a non-unitary answer. However the alternatives are highly
constrained by low-energy consistency. Two of the
alternatives involve low-energy effective non-locality, as advocated
for example in \sussc,
or a very long scattering time, as advocated for example in \jpas.
It is also very hard to imagine how either of these features could emerge
in a D-brane description. Our results of so far are consistent 
with all of these proposals, and do not tell us definitively 
how string theory 
solves the information puzzle. Nevertheless we have more clues 
and are optimistic that 
further progress on this issue is now possible.

We could consider other compactifications--for example heterotic
string compactified on $T^6$ ($N=4$) or on  $K3\times T^2$ ($N=2$),
which are dual
to type II strings on $K3\times T^2$ or on Calabi-Yau. The
Bekenstein-Hawking entropy for the $N=4$ cases have been
computed in \refs{\gmrn,\kall,\cvet}, while $N=2$ cases appear in \fks.
It is not too difficult in these examples
to set up the computation for the BPS states which preserve only
one unit of supersymmetry.  In one formulation, it is related
to the study of cohomology of
 moduli space of stable holomorphic $SU(N)$ bundles
on the six-manifold with fixed second and third Chern classes
determined by the charges.
Unfortunately at the present the dimension of the cohomology of such
moduli spaces is not known.  It would be interesting to compute
the growth of the cohomology of the moduli spaces for these bundles on
the six-manifolds and check the prediction obtained from the
Bekenstein-Hawking entropy.

\centerline{\bf Acknowledgements}

We would like to thank G. Horowitz, J. Polchinski,
S. Shenker and E. Witten for valuable discussions.
C.V. also thanks the hospitality of Rutgers University where this
work was completed.
The research of C.V. is supported in part by NSF grant PHY-92-18167.
The research of A.S. is supported in part by DOE grant DOE-91ER40618.

\listrefs
\end